\documentstyle[preprint,prl,aps,epsfig]{revtex}
\begin{document}
\title{Folding Pathways of Prion and Doppel}
\author{Gianni Settanni$^*$, Trinh Xuan Hoang$^{*,\dagger}$, 
Cristian Micheletti$^*$ and Amos Maritan$^{*,\dagger,**}$}
\vskip 0.3cm
\address{$^*$International School for Advanced Studies (S.I.S.S.A.)
and INFM, Via Beirut 2-4, 34014 Trieste, Italy;
$^{\dagger}$The Abdus Salam International Center for Theoretical
Physics (ICTP), Strada Costiera 11, 34100 Trieste, Italy}
\date{\today}
\maketitle

\tighten

\vskip 1.cm
\noindent
$^{**}${\em Corresponding author:} \\
Amos Maritan \\
International School for Advanced Studies (SISSA/ISAS) \\
via Beirut 2-4 \\
34014 Trieste \\
ITALY \\
tel: +39.040.2240462 \\
fax: +39.040.3787528 \\
email: maritan@sissa.it




\newpage
\vskip 0.5cm
\begin{abstract}
The relevance of various residue positions for the stability and the
folding characteristics of the prion protein are investigated by using
molecular dynamics simulations of models exploiting the topology of
the native state.  Highly significant correlations are found between
the most relevant sites in our analysis and the single point mutations
known to be associated with the arousal of the genetic forms of prion
disease (caused by the conformational change from the cellular to the
scrapie isoform). Considerable insight into the conformational change
is provided by comparing the folding process of prion and doppel (a
newly discovered protein) sharing very similar native state topology:
the folding pathways of the former can be grouped in two main classes
according to which tertiary structure contacts are formed first
enroute to the native state. For the latter a single class of pathways
leads to the native state. Our results are consistent and supportive
of the recent experimental findings that doppel lacks the scrapie
isoform and that such remarkably different behavior results from
differences in the region containing the two $\beta-$strands and the
intervening helix.
\end{abstract}

\section{Introduction}

Neurodegenerative diseases causing transmissible spongiform
encephalopathies (TSE) are the subject of intense research. They
affect humans and animals and include scrapie in sheep, mad-cow
disease in bovines and Creutzfeldt-Jacob disease in humans. Most of
them arise sporadically, the rest of the instances are inherited or
transmitted by inoculation through the dietary or infected tissues.

The belief is that the normal, and benign, form of the prion protein
$PrP^c$, which is found in various organs of vertebrates, including
the brain, can undergo a post translational process leading to a
conformational change of its native state \cite{prusiner1}. 
This new form is designated
$PrP^{sc}$ and contrarily to $PrP^c$ it is insoluble. The NMR
investigation of the three-dimensional structure of recombinant
$PrP^{c}$ of various species\cite{3,4,5,6,10,11,12,13,14} has revealed
that the N-terminus is unstructured whereas the C-terminus consists of
three $\alpha$-helices and two short $\beta$-strands.

On the other hand the structure of the malign form, $PrP^{sc}$, is
mostly unknown. However, spectroscopic investigations, mainly based on
circular dichroism, showed that $PrP^c$ contains about 40\%
$\alpha$-helices and little $\beta$-sheet, whereas$PrP^{sc}$ is
composed of about 30\% $\alpha$-helices and 45\% $\beta$-sheet
\cite{15,16}.

Unfolding and re-folding experiments suggest that the conversion
between the $\alpha$-rich and $\beta$-rich form occurs through a
complete unfolding of the protein\cite{17}.  Experiments and clinical
tests suggest two main processes for the onset and spreading of the
prion infection. In the first case, the spreading of the scrapie form
in healthy tissues contaminated with $PrP^{sc}$ reveals that the
presence of $PrP^{sc}$ helps further conversion from normal to scrapie
form, i.e. $PrP^{sc}$ acts as a template for the restructuration of
$PrP^c$. In the second case, the genetic influence on the propensity
of cellular $PrP^c$ to rearrange into $PrP^{sc}$ is revealed by the
discovery of, at least, 20 single point mutations in the PrP gene in
Humans which favor the spontaneous onset of the disease
\cite{5,6,7,8,9}.

Within the picture presented in \cite{17} the role of the mutations in
the arousal of the disease could be caused by several factors ranging
from the lowering of a free energy barrier in the conversion process to
the increase in the oligomerization rate. One of the goals of the
present study is to understand which are the sites where mutations are
expected to have the major impact in the stabilization/destabilization
of the $PrP^c$ structure and/or its folding process.  

Recently, some experiments on transgenic mice lacking the $PrP$ gene
($PrnP^{0/0}$) have revealed the onset of neurodegeneration
(Purkinje-cell death) \cite{24} completely different from usual prion
disease. This disease was ultimately traced back to a prion homologous
gene, $Prnd$, (not expressed in normal subjects), which encoded a
protein named doppel, $Dpl$ which has almost the same topology of
$PrP^c$, despite the low sequence identity (25\%). 

A striking difference between $Dpl$ and $PrP^{sc}$, is that the former
affects the central nervous system while retaining its native structure,
i.e. without a need to convert to a scrapie-like conformation. 
Various hypothesis have been formulated \cite{26} to explain these
different behaviours.

The theoretical framework we elaborated and adopted in our study helps
to shed light on the putative causes for such remarkably different
behavior of doppel and prion. Indeed, the different folding behavior
and stability of prion and doppel are related, at least in part, to
the small, but important, differences in the topology of the native
state which affect in an amplified fashion, the folding routes and
ultimately facilitating structural rearrangements of prion.

The model we adopt builds on the importance of the native state
topology in steering the folding process, that is in bringing into
contact pairs of amino acids that are found in interaction in the
native state. In the past few years an increasing amount of
experimental \cite{27,28,29} and theoretical
\cite{dos,fast,nature,baker,eaton,alm,finkelstein,cecilia,cris_prep}
evidence (such as all-atom molecular dynamics (MD) simulations in
implicit solvent) have confirmed this view \cite{30,31}.

A further confirmation of this view is
provided by the remarkable accord of the key folding stages predicted by
topology-based models and the available experimental results
\cite{32,33}. Indeed, the success of the topological picture itself
helps to explain why the folding process is not too sensitive to the
detailed chemical composition of most residues in a protein. The
dependence of the folding process on the detailed chemistry is much more
subtle than that given by the native state topology.

Usually the folding mechanism is affected only when mutations occurs in
a small set of key residues. 
Within our theoretical framework those key residues take part to
contacts that are crucial for the folding process \cite{32}.

The establishment of the key contacts leads to a rapid formation of
further interactions. Interestingly, the folding bottlenecks can be
identified by just knowing the topology of the native conformation
\cite{32,33}. Thus, the topology itself also dictates the impact of
chemistry on the folding process.

The purpose of the present work is three-fold:

i) determination of the folding bottlenecks for the cellular prion and
the identification of the key amino acids taking part to the
corresponding crucial contacts;

ii) making connection between the set of such key residues with those
that are known to be associated with harmful $PrP^{sc}$ mutations. As
argued before, a random mutation on key positions will usually result
in a disruption of the folding process. Only fine-tuned mutations can
lead to a wild-type-like native state (as for viral enzymes mutating
under drug attack \cite {32}) or in another viable structure (as
postulated by Prusiner for the prion \cite{prusiner1});

iii) furthermore, we will focus our attention on how the topological
differences between the native states of $Dpl$ and $PrP^c$ may have
impact on the folding process thus aiding or avoiding the formation of
misfolded conformers.  It is important to stress that our study is
based on the native-state structures of $PrP^c$ and $Dpl$. Hence,
although we can confidently identify the crucial folding residues, we
cannot confirm explicitly that their mutation results in a different
native state form.

\section{Materials and Methods}
\subsection{The model for prion}

Our model builds on a schematic representation of the
three-dimensional structure of the protein, where the amino acids are
replaced by effective centroids identified with the $C_{\alpha}$
atoms. The bias towards the native state is introduced through a
Go-like \cite{40} energy scoring function that rewards the formation
of native contacts between the centroids. The list of native
interactions was compiled from the knowledge of the coordinates of
atoms of the human prion protein (PDB code 1qlx). A pair of amino
acids is considered in interaction if any pair of their heavy atoms,
$i$ and $j$, have a native separation smaller than the distance,
$1.244 (R_i + R_j)$, where the point of inflection of the Van der
Waals interaction occurs ($R_i$ denotes the Van der Waals
radius of atom $i$). The values of $R_i$ are taken from Ref. \cite{36}.
In this way one defines a symmetric matrix, known
as contact map, $\Delta$, whose entries, $\Delta_{ij}$ are equal to 1
if the $i$-th and $j$-th centroids interact, and zero otherwise.

The energy function for our system includes terms that are routinely
used in standard molecular dynamics (MD) simulations on biopolymers
\cite{37,38}. It is composed by a "bonded" and a
"non-bonded" term:
$E=V_B + V_{NB}$. The former accounts for the constraints such as the
peptide bond length and Ramachandran angles bias acting on the
aminoacids at a local level. Its explicit form is:
\begin{eqnarray} \label{VB}
V_B&=&g\cdot V_p+h\cdot V_a+k\cdot V_d \; ,\\
V_p&=& \sigma \sum_{i=1}^{N-1} \left(r_{i,i+1}-r_{i,i+1}^{(n)}\right)^2 \; , \\
V_a&=&\sigma \sum_{i=1}^{N-2}
\left(\theta_{i,i+1,i+2}-\theta_{i,i+1,i+2}^{(n)}\right)^2 \; , \\
V_d&=&\sigma \sum_{i=1}^{N-3} \left(1
-\cos\left(\tau_{i,i+1,i+2,i+3}-\tau_{i,i+1,i+2,i+3}^{(n)}\right)\right) \; ,
\end{eqnarray}
where $r_{i,j}$ is the distance between residue $i$ and $j$,
$\theta_{i,j,k}$ is the angle with the $j$-th amino acid as vertex
and the $i$-th and the $k$-th as edges and $\tau_{i,j,k,l}$ is the
dihedral generated by the $i$-th, the $j$-th, the $k$-th and the
$l$-th amino acid. The $n$ as superscript denotes the native state
value. $\sigma$ represents a suitable scale factor to fix the
temperature scale (set to $10$ in our simulations) which is given in
dimensionless units.

The minimum of expression ($\ref{VB}$) is precisely attained
in correspondence of the native state. However the cooperative
character of the folding process \cite{Kaya} can be captured only by introducing
an explicit bias towards the formation of native interactions
\cite{gaussian}. This is accomplished through the second term,
$V_{NB}$, which weights the interaction of any pair of non-consecutive
centroids with a Van der Waals-like potential:

\begin{equation} \label{VdW}
V_{NB}=\sigma \sum_{i<j-3}^N V_{ij} = \sigma \sum_{i<j-3}^N
\left(5\left(\frac{r_{ij}}{r_{ij}^{(n)}}\right)^{-12}- \;\;
6\left(\frac{r_{ij}}{r_{ij}^{(n)}}\right)^{-10}\Delta_{ij}\right) \; .
\end{equation}

\noindent
This choice rewards native interactions and disfavors
non-native ones. The form of the energy scoring function ensures that
the global minimum is attained in correspondence of the native state,
regardless of the precise values for the (positive) couplings
parameters $g$, $h$, $k$. The values used in the present study are: $g
= 50$, $h = 5$ and $k = 0.3$; we have checked that the results are
robust to their variation within a certain range \cite{33}.

The dimensionless unit of time in our MD simulations is $5\cdot
10^{-3}$ and conformations were sampled every $500$ time steps to allow a
sufficient uncorrelation at temperatures below the folding transition.
The multiple-histogram technique \cite{41} has been applied to
reweight quantities such as the average internal energy collected in
more than 50 equilibrated runs at different temperatures in the range
2-8. The optimal reweighting allows to calculate thermodynamic
quantities, such as the average internal energy and the specific heat
for an arbitrary temperature, $T$.

As visible in Fig. \ref{fig:fig2}A, the specific heat exhibits a
single peak which signals the folding transition \cite{hao}. In
addition to this overall characterization of the equilibrium
thermodynamics, it is possible to monitor how the individual
interactions that are present in the native state are formed as the
folding progresses. Indeed, although all native contacts are
energetically favored in the same way, their entropic cost of
formation may significantly vary according to their locality,
burial/exposure etc. .

The crucial contacts for the folding process are identified according
to the method given in Refs. \cite{32,33,gaussian}, which singles out the
contacts giving the largest contribution to the overall specific heat
at the folding transition temperature, $T_F$.

\begin{equation} \label{cij}
c_{ij} = \frac{d \left< V_{ij} \right>}{d T} = \left(\left< V_{NB}V_{ij} \right> -
\left<V_{NB}\right>\left<V_{ij}\right>\right) / kT^2
\label{cv}
\end{equation}

The physical interpretation of this procedure is that these special
contacts are precisely those that, having a significant and sudden
formation at $T_F$, act as bottlenecks for the folding process.
$c_{i,j}$ measures the sensitivity of the average energy of the system
to perturbation occurring at the contact $(i,j)$.

In our studies, a long run carried out at $T_F$ of $2\cdot 10^7$ time
steps, allowed a detailed analysis for all contacts. In addition to
the identification of crucial contacts, it is useful to introduce a
similar measure for the individual amino acids through the single
residue specific heat (SRSH) resulting from the contribution of all
contacts to which a given site, $i$, takes part to:

\begin{equation} \label{ci}
c_{i}=\sum_j \Delta_{ij} c_{ij}
\end{equation}

SRSH's have been used to identify folding and unfolding phases and to
pinpoint residues that play a relevant role in the process
\cite{32,33}. The crucial folding steps can thus be identified with two
(related) criteria: namely either through the contacts or the sites
with the largest specific heat of formation.

\subsection{Folding pathways}

Another issue that we have addressed in our study is to establish the
presence of alternative pathways taking from unfolded conformations to
the native state. Again, this is accomplished by examining in detail
both the thermodynamic and the kinetic relevant features extracted
from the model. The comparison of dynamic trajectories connecting
unfolded and native-like conformations has revealed the presence of
several folding events that have to occur for the protein to fold.
The main ones that emerged from our analysis, and that are
particularly apt for characterizing the folding process, correspond to
the formation of contacts involving three sets of contiguous
residues. The first set, $Bh$, from residue 125 to residue 172 comprises
all N-terminal residues ($\beta$-sheet included) before the helix 2
(defined according to the indications in the PDB file). The second
set, $h1$, from residue 173 to 194 comprises residues of helix 2, and
finally the third set, $h2$, from residue 200 to 228 is composed of
residues from helix 3.

The four observed events are:

\begin{itemize}
\item{Event $Bh$}: formation of contacts between residues within $Bh$,
\item{Event $Bh$-$h1$}: formation of contacts between $Bh$ and $h1$,
\item{Event $Bh$-$h2$}: formation of contacts between $Bh$ and $h2$,
\item{Event $h1$-$h2$}: formation of contacts between $h1$ and $h2$.
\end{itemize}

For convenience of presentation, in order to describe the four events
on a similar footing, we focus on the top 10 contacts (in terms of
specific heat contributions $c_{ij}$) in each interacting group.
It is interesting to note
that, among this top contacts, those occurring within an $\alpha$
helix almost never occur. This can be explained since local contacts,
among others, are easily formed, even at high temperature, and hence
do not constitute a bottleneck for overcoming the configurational
barrier to the native state.

We computed the fraction of formed contacts, $Q_i$ ($i$ = $Bh$, $Bh$-$h1$,
$Bh$-$h2$, $h1$-$h2$), within each of the 4 groups along the dynamical
trajectories of $2\cdot 10^7$ steps at $T_F$. The natural binning size
for each of the $Q_i$ was $1/10$. From this, we obtained a vivid
picture of the dynamical effects of the free energy landscape at the
folding transition (see Fig. \ref{fig:fig7} and the Results section).

We also carried out 100 (``quenched'') folding runs at
$T=\left(1-\epsilon \right)T_F$, where $\epsilon = 1/30$, slightly
below $T_F$, starting from completely unfolded conformations
(thermalised at high temperature runs). In all cases, the native state
was reached in less than $2\cdot 10^5$ steps. The average $Q_i$'s as
a function of the elapsed time using all $100$ trajectories was also
calculated (Fig. \ref{fig:fig7} ).

\subsection{Doppel protein}

We carried out a parallel analysis for the doppel protein, $Dpl$, whose
NMR-resolved structure was retrieved from the PDB (PDB code 1i17, model
1). For this protein too, we built the contact map considering
all-atom distances, and obtained both the overall
specific heat (Fig. \ref{fig:fig2}B ) and the one associated to individual contacts using the
same methods described above. A crucial difference in the folding
events of the two proteins was revealed by monitoring the four
events, analogous to those of the previous section.

For $Dpl$ the sets of amino acids structurally homologous of the prion
ones are as follows: the set $Bh$ corresponds to residues from 1 to 50
before helix 3; $h1$ to residues from 51 to 76 (comprising helix 3 and
4) and the set $h2$ to residues from 78 to 96 (comprising helix 5). The
numeration of the helices is consistent with the secondary structure
indications present in the PDB file. Free energy profiles and time
dependent averages of the reaction coordinate are shown in Fig.
\ref{fig:fig6} and \ref{fig:fig7} respectively.

\section{Results}

\subsection{Determination of key residues}

The specific heat curves as a function of the temperature for $PrP^c$
and $Dpl$ are shown in Figures \ref{fig:fig2}A and and \ref{fig:fig2}B,
respectively. The single peak in each of them allows to identify
unambiguously the folding transition in these systems.

From the long run of $2\cdot 10^7$ time steps performed at $T_F$, the
SRSH's for prion have been computed according to eq.(\ref{ci}). We
have ranked the residues according to their SRSH (see Table 1). It is
natural to compare the top key sites from our analysis with those
whose mutation is known to favor the emergence of $PrP^{sc}$
(highlighted in Table 2). In the top 10 amino acids of our list we
found 4 of the known key mutating sites; whereas in the top 30 amino
acids their number increases to 12. The statistical relevance of this
match can be obtained from a combinatorics calculation of the
probability to have at least the observed number of matches by pure
chance. For the top 10 amino acids the probability to get a better
results by chance is $6.8\%$ whereas for the top 30 amino acids that
probability becomes $0.02\%$. Thus, the criterion based on the SRSH
allows to identify confidently amino acids important for the folding
process of the prion protein.

Their structural location and role is made apparent in Fig.
\ref{fig:fig4} where the prion contact map has been colored according
to the value of the SRSH. As visible, crucial contacts connect
secondary-structure elements thus forming the protein tertiary
structure. In particular they connect the $\beta-$sheet-helix-1 region
to the helix-2 and helix-3 ($Bh$-$h1$ and $Bh$-$h2$ sets, respectively) and
helix-2 - helix-3 ($h1$-$h2$ set). The parts of the protein contributing
more to the specific heat, within our model, are located at the
C-terminal strand of the $\beta$-sheet, and at the two C-terminal
helices, especially helix-3 (see Fig. \ref{fig:fig5}).

\subsection{Folding pathways}\label{fp}

The analysis of the long run of $2\cdot 10^7$ time steps carried out
at $T_F$ has led to the construction of an {\em effective} free energy
landscape. It is based on the reaction coordinates $Q_i$ as defined in
the Methods section. The free energy as a function of pairs $(Q_i,
Q_j)$ for all six combinations of $i$ and $j$ have been calculated as

\begin{equation} \label{f}
F\left(Q_i,Q_j\right ) = -\ \ln\ \left(P\left(Q_i,Q_j\right)\right)
\end{equation}

\noindent where $P(Q_i,Q_j)$ is the fraction of times the dynamical
trajectory spends in the bin around $(Q_i,Q_j)$. The contour plots of
$F$ are shown in Figures \ref{fig:fig6}A and \ref{fig:fig6}B for
$PrP^c$ and $Dpl$ respectively. The two energy minima around $Q = 0$
and in the higher $Q$ region correspond to the unfolded and
(partially) folded conformations respectively. For prion the plots
involving set $h1$-$h2$ show the presence of multiple pathways
connecting these two minima. For example, the contour plot in the
bidimensional space spanned by $Q_{h1-h2}$ and $Q_{Bh-h1}$ presents
two possible successions of events: either the $Bh$-$h1$ set is formed
followed by formation of $h1$-$h2$ set or {\em vice versa} (see
Fig. \ref{fig:fig6}A). The same observation can be done for the other plots
involving set $h1$-$h2$. Such pathway ambiguity is absent for the case
of doppel (Fig. \ref{fig:fig6}B), where trajectories can be grouped
essentially in a single set. Indeed, for doppel, on average, the
formation of the $h1$-$h2$ contacts follows the other sets. This result
is further confirmed by the analysis of the time-dependent average of
the reaction coordinates $Q_i$'s as obtained from the $100$
non-equilibrium folding runs. Figures \ref{fig:fig7}A and
\ref{fig:fig7}B show the results for $PrP^c$ and $Dpl$
respectively. For $Dpl$ the order in the formation of the contacts is
the same detected by the the equilibrium run: set $Bh$ is formed
before set $Bh$-$h1$ that is followed by $Bh$-$h2$ eventually followed by
$h1$-$h2$ interhelical contacts. On the other hand, such definite
succession of events is not observed for prion.

\section{Discussion}

The goal of our numerical study was to characterize the folding
process of the prion protein elucidating the structural mechanisms
responsible for the propensity of $PrP^c$ to mutate into the
harmful scrapie form, $PrP^{sc}$.

At the heart of our analysis is a topology-based energy function
that, by rewarding the formation of native-like interactions
among the residues, allows to mimic the progressive build-up of native
structure in a folding process. Remarkably, the progress towards the
native state does not advance smoothly, but occurs through the
overcoming of configurational barriers, which are responsible for the
appearance of a marked peak in the specific heat curve. Crucial
contacts (and sites) for the folding process have thus been identified
as those giving the largest contributions to the specific heat peak
(Fig. \ref{fig:fig4}).

The top sites isolated by our procedure contain a highly significant
fraction of residues that are known to enhance the misfolding
propensity of $PrP^c$. This indicates that there is little freedom in
choosing the chemical identity of the amino acids at the sites which
have a special structural role during the folding
process.

The analysis presented here, based on topological arguments, provides a
physically-appealing interpretation for the connection between certain
site mutations and the arousal of the prion disease. Conversely, the
key mutating sites not captured by our analysis are presumably crucial
because they affect some aspects of the process leading to the scrapie
form that cannot be accounted for by the present method. For example,
the alternative mechanism leading to misconformation is the template
binding of $PrP^c$ to $PrP^{sc}$ involving the interconversion of the
former into the latter one. Consistently with this interpretation,
amino acids that do not present high SRSH and that are known to be
involved in the development of the disease, specifically MET 129,
GLU200 and GLU 219, (i.e. mostly charged residues with a large exposed
surface area).

\begin{figure}
\centerline{\epsfig{figure=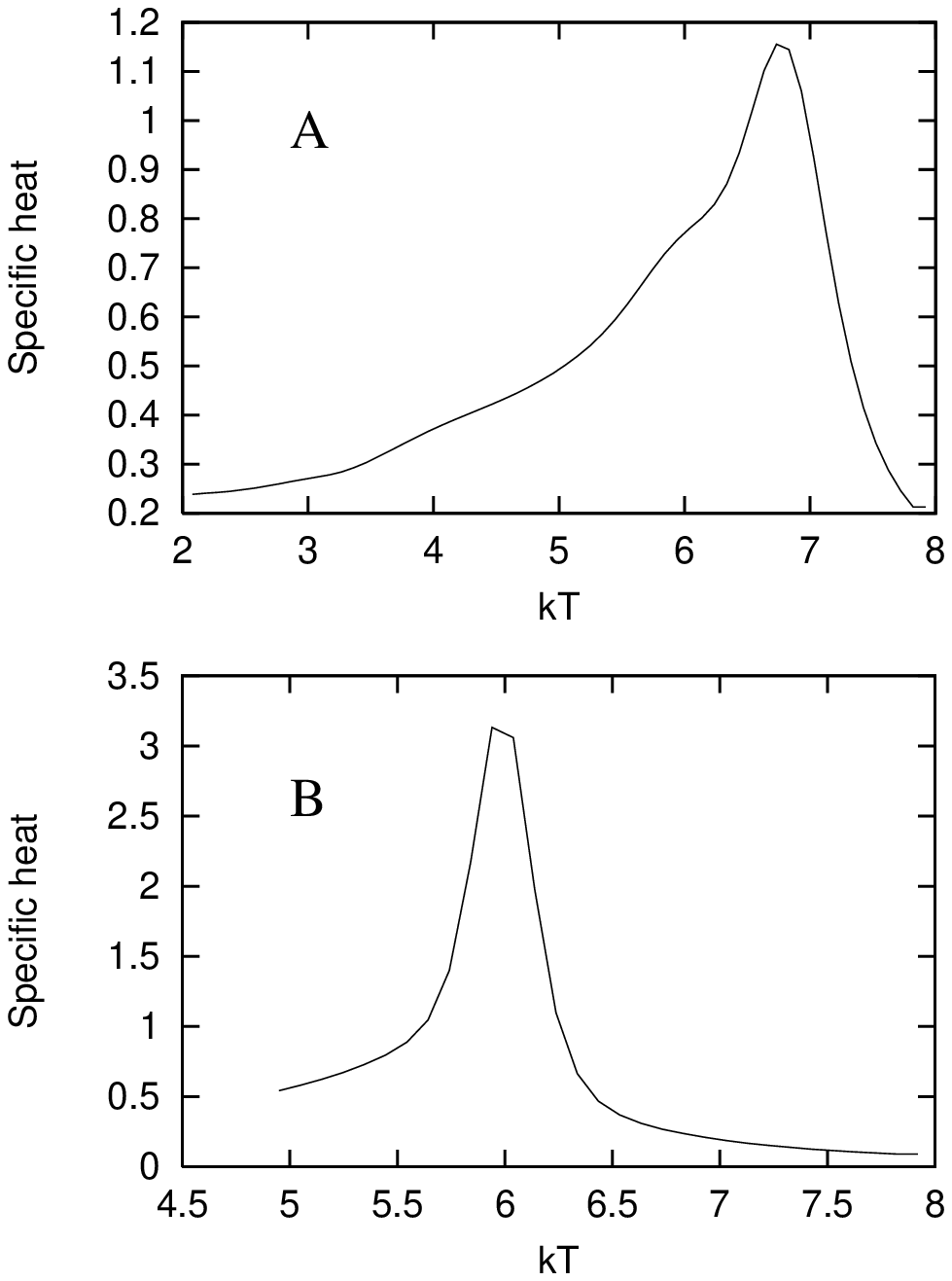,width=3.4in}}
\caption{}
\label{fig:fig2}
\end{figure}

\begin{figure}
\centerline{\epsfig{figure=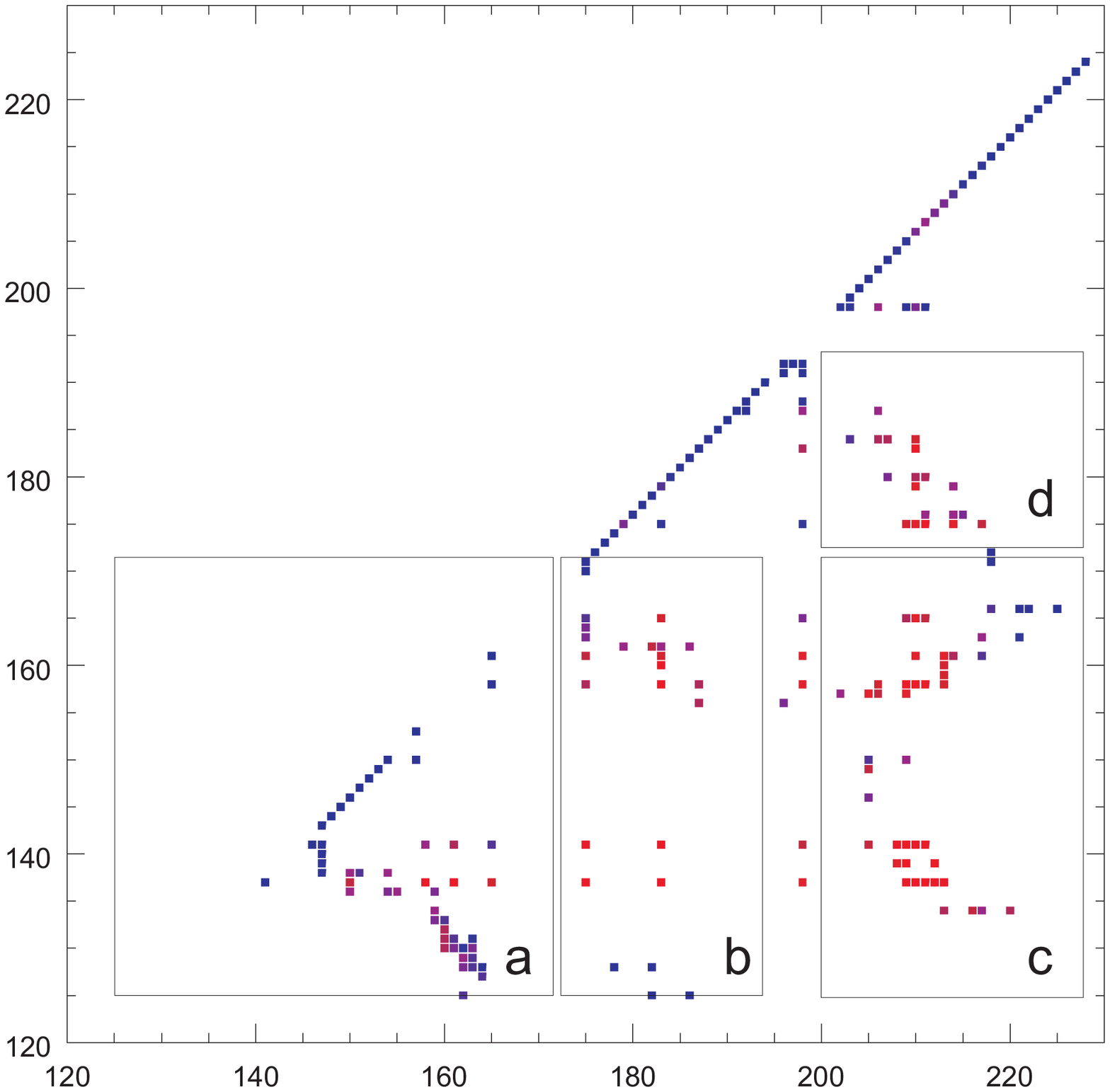,width=3.4in}}
\caption{}
\label{fig:fig4}
\end{figure}

\begin{figure}
\centerline{\epsfig{figure=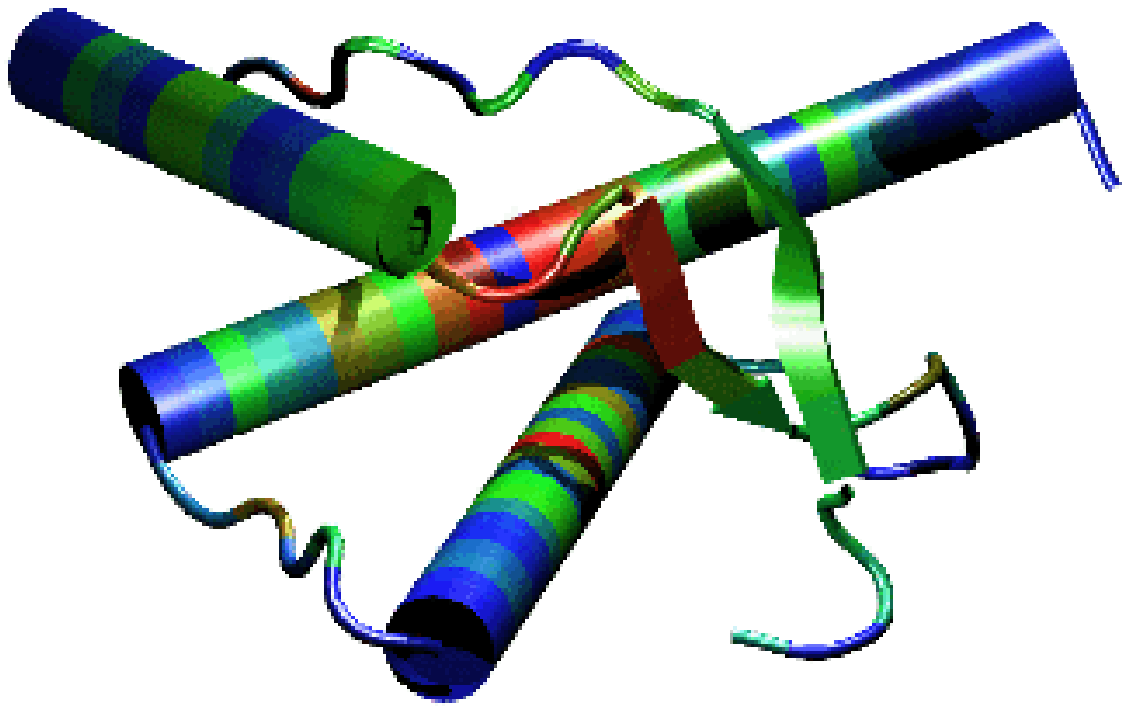,width=3.4in}}
\caption{}
\label{fig:fig5}
\end{figure}

\begin{figure}
\centerline{\epsfig{figure=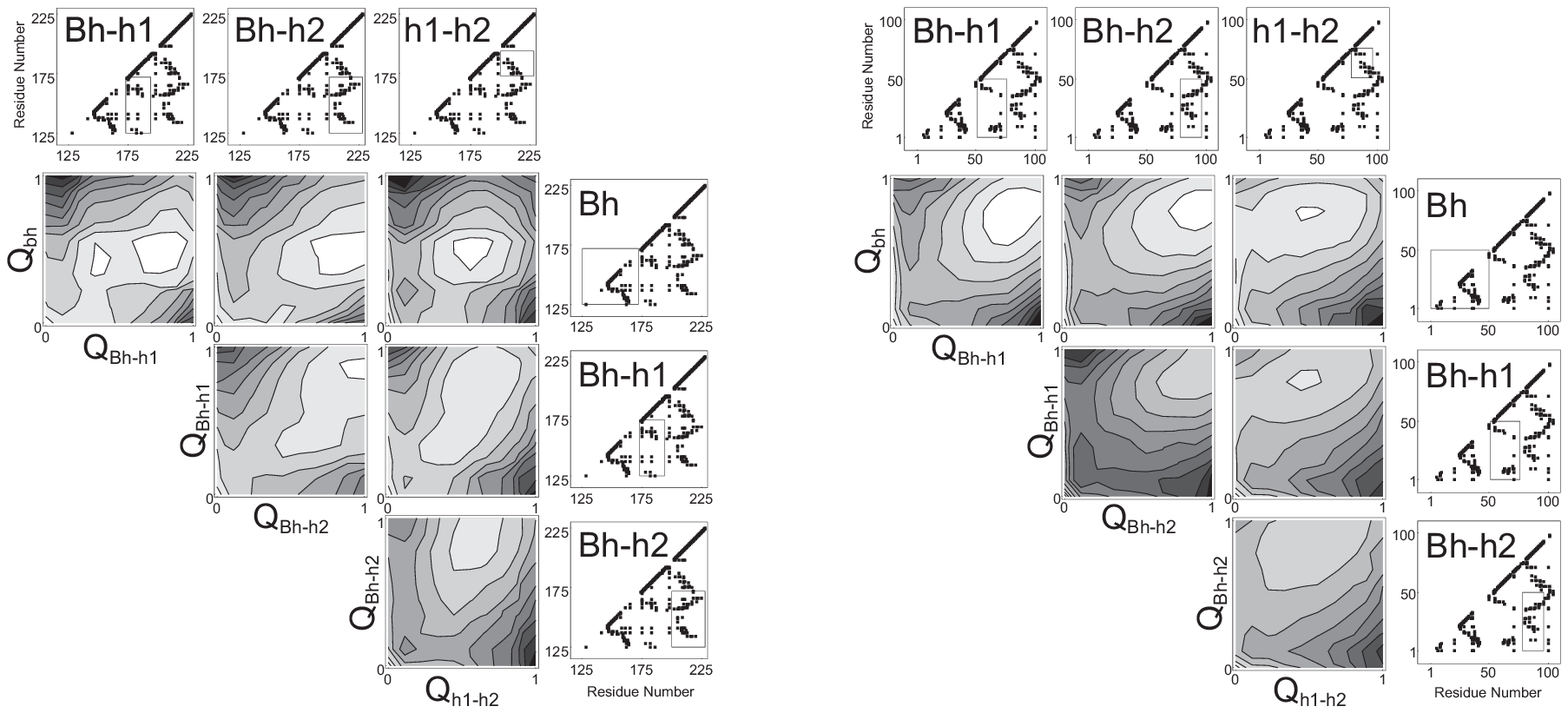,width=6.in}}
\caption{}
\label{fig:fig6}
\end{figure}

\begin{figure}
\centerline{\epsfig{figure=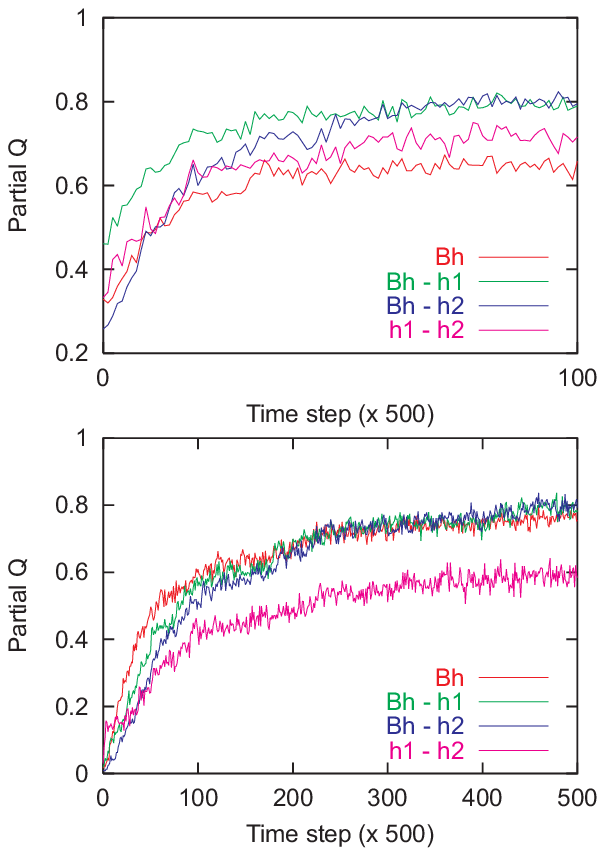,width=3.4in}}
\caption{}
\label{fig:fig7}
\end{figure}

Despite the substantial topological similarities of $Dpl$ and $PrP^c$,
the detailed analysis of the folding trajectories of both proteins
have revealed significant differences of the folding pathways.
Strikingly, for $PrP^c$ the folding pathways can be classified in two
main distinct routes, whereas for $Dpl$ only one of these is
essentially present. For the latter, the series of folding steps is
more distinctly marked than for the former. Such noticeably different
behavior can be ascribed to the different set of contacts stabilizing
the assembly of the $\alpha$-helices and $\beta$-sheets in the two
proteins.  It is interesting to note that the contact order \cite{29}
of $PrP^c$ is 0.2056 and is lower than that of $Dpl$ which is 0.2298,
even though the former has somewhat higher number of contacts which
is, in turn, reflected by the shift of the folding transition peak
towards higher temperatures (see Fig. \ref{fig:fig2}). This
difference could explain the different average folding time measured
in our models for the two proteins. Indeed, according to
Fig. \ref{fig:fig7}, the sets of contacts in prion are all formed
within the first 25000 time steps while contacts in doppel contacts
can take 4-5 times longer to form. This difference is of the same
order of magnitude of that predicted by Plaxco \cite{29} on the basis
of the contact order of the two proteins. Although the present picture
would certainly be influenced by the introduction of amino acid
specific interactions, it is appealing to connect the different
folding routes to the appearance of misfolded conformers. According to
this view, the latter would result from following the folding route of
$PrP^c$ where the inter C-terminal helices contacts (set $h1$-$h2$)
are formed while the contacts with the $\beta$-sheet (set $Bh$-$h1$
and $Bh$-$h2$) are not yet formed (Figures \ref{fig:fig6}A and
\ref{fig:fig7}A). Indeed, along this route, which is alternative to
the one where the two events are interchanged as it occurs in the
single route of doppel (Fig.  \ref{fig:fig6}B and \ref{fig:fig7}B),
the N-terminal part of the protein is free to rearrange its structure
before reaching the native-state. This scenario is consistent with
several recent models (see e.g. Ref. \cite{Peretz,Huang}) that
identify the N-terminal part of the protein as the one that undergoes
the conformational change that leads the scrapie form of prion.

\newpage

\begin{table}
\begin{tabular}{|cc|cc|cc|}
rank & site & rank & site & rank & site\\ \hline
1  & 137 & 11 & 213 & 21 & 214 \\
2  & 210 & 12 & 165 & 22 & 163 \\
3  & 175 & 13 & 160 & 23 & 212 \\
4  & 158 & 14 & 162 & 24 & 184 \\
5  & 141 & 15 & 206 & 25 & 187 \\
6  & 183 & 16 & 205 & 26 & 179 \\
7  & 209 & 17 & 157 & 27 & 208 \\
8  & 198 & 18 & 139 & 28 & 217 \\
9  & 161 & 19 & 150 & 29 & 159 \\
10 & 211 & 20 & 134 & 30 & 180 \\
\end{tabular}
\label{tab:tab1}
\vskip 1.0cm
\caption{Top 30 sites in prion ranked according to the single residue specific
heat (SRSH).}

\end{table}


\begin{table}
\begin{tabular}{|c|}
\hline
145, \underline{160}, 178, \underline{179}, \underline{180},
 \underline{{\bf 183}},
\underline{187}, 188, 196,\\ \underline{{\bf 198}},
 200, 202, \underline{208}, \underline{{\bf 210}}, \underline{{\bf 211}},
 \underline{212}, \underline{214}, \underline{217}\\
\hline
\end{tabular}
\label{tab:tab2}
\vskip 1.0cm
\caption{Sites where mutations (in the NMR-solved part of Human $PrP$)
have been observed to cause/stop the prion disease or be fundamental
for correct folding of $PrP^c$. Sites 214 and 179 represent cysteine
residues that form a disulfide bond needed to observe the $PrP^c$
$\alpha-$form of prion [32].  Single residue specific
heat (SRSH), as determined by our model, ranking within the top 10 are
in boldface; those ranking within the top 30 are underlined (see Table
1).}

\end{table}

\newpage

\centerline{\Large Figure captions}

\begin{enumerate}

\item{Figure 1.} Temperature dependent specific heat for the prion (A)
and doppel (B) models. The folding transition temperature is
identified at the point where the specific heat peak occurs.

\item{Figure 2.} Contact map of $PrP^c$ (PDB code 1qlx) where contacts are
highlighted according to their specific heat $c_{ij}$ of equation
(\ref{cv}). Red spots correspond to contacts with high $c_{ij}$,
while blue ones correspond low SRSH. The relevant sets of contacts (as
defined in the text) are framed within the rectangles; (a) set $Bh$, (b)
set $Bh$-$h1$, (c) set $Bh$-$h2$, (d) set $h1$-$h2$. Notice that red spots are
prevalent in all sets but $Bh$.

\item{Figure 3.} The three-dimensional structure of prion protein
where amino acids have been colored according to the value of their
SRSH. Red spots are located mainly on the C-terminal strand of the
$\beta-$sheet, on helix-2 and helix-3.

\item{Figure 4.} Effective free energy landscape for prion (A) and
doppel (B) model. Upper and right squares represent the contact maps
where the representative set of contacts defining the reaction
coordinates are highlighted (see Method section). Contour plots are
shown for each of the six pairs of reaction coordinates
$(Q_i,Q_j)$. White (black) spots represent low (high) values of the
free energy. The minimum close to (0,0) corresponds to the unfolded
state whereas the minimum closer to (1,1) corresponds to the folded
state. In most of the contour plots for prion (A) more than one class
of favorable paths from the unfolded to the folded conformation is
possible whereas for doppel (B) only one class of paths is the most
favorable. Following these pathways allows to assign a preferential
order to the events. Thus, on average, the sequence of contact
formation for doppel is $Bh$, $Bh$-$h1$, $Bh$-$h2$ and $h1$-$h2$ 
(see also Fig. \ref{fig:fig7}).
For prion, two class of equivalent pathways are present: in one
case the $h1$-$h2$ contacts are formed before the other sets of contacts;
in the other case $h1$-$h2$ contacts follow the formation of all the other
contacts as for $Dpl$.

\item{Figure 5.} Average time dependence of reactions coordinates
$Q_i$'s for prion (A) and doppel (B) computed on 100 folding runs
starting from completely unfolded conformations. (A) In case of prion
the curves corresponding to the different sets (red for $Bh$, green for
$Bh$-$h1$, blue for $Bh$-$h2$ and magenta for $h1$-$h2$) present
intersections. Indeed the average over the different folding pathways
results in a non well-defined sequence of folding events (B) The
folding events in the case of doppel protein have a well defined
sequence: $Bh$ set (red) is the first to form; then $Bh$-$h1$ (green);
then $Bh$-$h2$ (blue); finally $h1$-$h2$ (magenta).

\end{enumerate}

\newpage

\newpage

\begin{thebibliography}{99}

\bibitem{prusiner1}
Prusiner S.B. (1982) {\em Science}, {\bf 216}, 136-144.

\bibitem{3} Goldgaber, D., Goldfarb, L. G., Brown, P., Asher, D. M., Brown, W.
T., Lin, S.,  Teener, J. W., Feinstone, S. M., Rubenstein, R., Kascsak, R.J. et
al. (1989) {\em Exp.  Neurol.} {\bf 106}, 204-206.

\bibitem{4} Kretzschmar, H. A., Honold, G., Seitelberger, F., Feucht, M.,
Wessely, P.,  Mehraein, P. \& Budka, H. (1991) {\em Lancet} {\bf 337},
1160.

\bibitem{5} Hsiao, K., Baker, H. F., Crow, T. J., Poulter, M., Owen, F.,
Terwilliger, J. D.,  Westaway, D., Ott, J. \& Prusiner, S. B. (1989)
{\em Nature} {\bf 338}, 342-345.

\bibitem{6} Gabizon, R., Rosenmann, H.,Meiner, Z., Kahana, I., Kahana, E.,
Shugart, Y., Ott,  J. \& Prusiner, S. B. (1993)
{\em Am. J. Hum. Genet.} {\bf 53}, 828-835.

\bibitem{10} Calzolai, L., Lysek, D. A., Guntert, P., von Schroetter, C.,
Riek, R., Zahn, R. \& Wuthrich, K. (2000)
{\em Proc. Natl. Acad. Sci. USA} {\bf 97}, 8340-834.

\bibitem{11} Lopez, G. F., Zahn, R., Riek, R. \& Wuthrich, K. (2000)
{\em Proc. Natl.  Acad. Sci. USA} {\bf 97}, 8334-8339.

\bibitem{12} Liu, A., Riek, R., Wider, G., von Schroetter, C., Zahn, R. \&
Wuthrich, K. (2000)
{\em J. Biomol. NMR}  {\bf 16}, 127-138.

\bibitem{13} Zahn, R., Liu, A., Luhrs, T., Riek, R., von Schroetter, C.,
Lopez, G. F., Billeter, M., Calzolai, L., Wider, G. \& Wuthrich, K. (2000)
{\em Proc.  Natl. Acad. Sci. USA} {\bf 97}, 145-150.

\bibitem{14} James, T. L., Liu, H., Ulyanov, N. B., Farr-Jones, S., Zhang,
H., Donne, D. G., Kaneko, K., Groth, D., Mehlhorn, I., Prusiner, S. B. et al.
(1997) {\em Proc. Natl. Acad. Sci. USA} {\bf 94}, 10086-10091.

\bibitem{15} Pan, K. M., Baldwin, M., Nguyen, J., Gasset, M., Serban, A.,
Groth, D., Mehlhorn, I., Huang, Z. W., Fletterick, R. J., Cohen, F. E.
et al. (1993)
{\em Proc. Natl. Acad. Sci. USA} {\bf 90}, 10962-10966.

\bibitem{16} Pergami, P., Jaffe, H. \& Safar, J. (1996)
{\em Anal. Biochem.} {\bf 236}, 63-73.

\bibitem{17} Baskakov, I. V., Legname, G., Prusiner, S. B. \& Cohen, F. E.
(2001) {\em J. Biol. Chem.} {\bf 276}, 19687-19690.

\bibitem{7} Dlouhy, S. R., Hsiao, K., Farlow, M. R., Foroud, T., Conneally,
P. M., Johnson, P., Prusiner, S. B., Hodes, M. E. \& Ghetti, B. (1992)
{\em Nat.  Genet.} {\bf 1}, 64-67.

\bibitem{8} Petersen, R. B., Tabaton, M., Berg, L., Schrank, B., Torack, R. M.,
Leal, S., Julien, J., Vital, C., Deleplanque, B., Pendlebury, W. W. et al.
(1992) {\em Neurology} {\bf 42}, 1859- 1863.

\bibitem{9} Poulter, M., Baker, H. F., Frith, C. D., Leach, M., Lofthouse, R.,
Ridley, R. M., Shah, T., Owen, F., Collinge, J., Brown, J. et al. (1992)
{\em Brain} {\bf 115}, 675- 685.

\bibitem{24} Sakaguchi, S., Katamine, S., Nishida, N., Moriuchi, R.,
Shigematsu, K., Sugimoto, T., Nakatani, A., Kataoka, Y., Houtani, T., Shirabe,
S. et al. (1996) Nature 380, 528-531.

\bibitem{26} Mo, H., Moore, R. C., Cohen, F. E., Westaway, D., Prusiner, S.
B., Wright, P. E. \& Dyson, H. J. (2001)
{\em Proc. Natl. Acad. Sci. USA} {\bf 98}, 2352-2357.

\bibitem{27} Riddle, D. S., Grantcharova, V. P., Santiago, J. V., Alm, E.,
Ruczinski, I., I \& Baker, D. (1999)
{\em Nat. Struct. Biol.} {\bf 6}, 1016-1024.

\bibitem{28} Martinez, J. C. \& Serrano, L. (1999)
{\em Nat. Struct. Biol.} {\bf 6}, 1010-1016.

\bibitem{29} Plaxco, K. W., Simons, K. T. \& Baker, D. (1998) J. Mol. Biol.
277, 985-994.

\bibitem{dos} 
Micheletti, C., Banavar, J. R., Maritan, A. \& Seno, F. (1999)
{\em Phys. Rev. Lett.} {\bf 82}, 3372-3375.

\bibitem{fast} 
Maritan, A., Micheletti, C. \& Banavar, J. R. (2000)
{\em Phys. Rev. Lett.} {\bf 84}, 3009-3012.

\bibitem{nature} 
Maritan, A., Micheletti, C., Trovato, A. \& Banavar, J. R. (2000)
{\em Nature} {\bf 406}, 287-290.

\bibitem{baker} 
Baker, D. (2000) {\em Nature} {\bf 405}, 39-42.

\bibitem{eaton}
Munoz, V. \& Eaton, W. A. (1999)
{\em Proc. Natl. Acad. Sci. USA} {\bf 96}, 11311-11316.

\bibitem{alm} 
Alm, E. \& Baker, D. (1999)
{\em Proc. Natl. Acad. Sci. USA} {\bf 96}, 11305-11310.

\bibitem{finkelstein} 
Galzitskaya, O. V. \& Finkelstein, A. V. (1999)
{\em Proc. Natl. Acad. Sci. USA} {\bf 96}, 11299-11304.

\bibitem{cecilia} 
Clementi, C., Nymeyer, H. \& Onuchic, J. N. (2000)
{\em J. Mol. Biol.} {\bf 298}, 937-953.

\bibitem{cris_prep} C. Micheletti, SISSA preprint (2001).

\bibitem{30} Ferrara, P. \& Caflisch, A. (2001)
{\em J. Mol. Biol.} {\bf 306}, 837-850.

\bibitem{31} Gsponer, J. \& Caflisch, A. (2001)
{\em J. Mol. Biol.} {\bf 309}, 285-298.

\bibitem{32} Cecconi, F., Micheletti, C., Carloni, P. \& Maritan, A. (2001)
{\em Proteins} {\bf 43}, 365-372.

\bibitem{33} Settanni G, Cattaneo A \& Maritan A, (2001)
{\em Biophys. Journ.} {\bf 80} 2935-2945


\bibitem{40} Go, N. (1983)
{\em Annu. Rev. Biophys. Bioeng.} {\bf 12}, 183-210.

\bibitem{36} Tsai, J., Taylor, R., Chothia, C. \& Gerstein, M. (1999)
{\em J. Mol.  Biol.} {\bf 290}, 253-266.

\bibitem{37} Pearlman, D. A., Case, D. A., Caldwell, J. W., Ross, W. S.,
Cheatham, T. E., Debolt, S., Ferguson, D., Seibel, G. \& Kollman, P. (1995)
{\em Computer Physics Communications} {\bf 91}, 1-41.

\bibitem{38} Brooks, B. R., Bruccoleri, R. E., Olafson, B. D., States, D. J.,
Swaminathan, S. \& Karplus, M. (1983)
{\em J. Comp. Chem.} {\bf 4}, 187-217.

\bibitem{Kaya} Kaya, H. \& Chan, H. S. (2000) {\em Phys. Rev. Lett} {\bf
85}, 4823-4826; Kaya, H. \& Chan, H. S. (2000) {\em Proteins} {\bf 40},
637-661.

\bibitem{gaussian} 
Micheletti, C., Banavar, J. R. \& Maritan, A. (2001)
{\em Phys. Rev. Lett.} {\bf 87}, art. no. 088102

\bibitem{41} Ferrenberg, A. M. \& Swendsen, R. H. (1989)
{\em Phys. Rev. Lett.} {\bf 63}, 1195-1198.

\bibitem{hao} 
Hao, M. H. \& Scheraga, H. A. (1997)
{\em Physica A} {\bf 244}, 124-146.


\bibitem{Peretz}
Peretz, D., Williamson, R. A., Matsunaga, Y., Serban, H., Pinilla,
C., Bastidas, R., Rozenshteyn, R., James, T. L., Houghten, R. A.,
Cohen, F. E., {\em et al} (1997) {\em J. Mol. Biol. }
{\bf 273}, 614-622.
 
\bibitem{Huang}
Huang, Z., Prusiner, S. B. \& Cohen, F. E. (1996) {\em Fold. Des.}
{\bf 1}, 13-19.

\end{thebibliography}
\end{document}